\documentclass[twocolumn,showpacs,preprintnumbers,amsmath,amssymb]{revtex4}

\usepackage{graphicx}
\usepackage{dcolumn}
\usepackage{bm}

\begin{document}

\title{Evidence for Radio Detection of Extensive Air Showers\\
Induced by Ultra High Energy Cosmic Rays}

\author{D. Ardouin}
\author{A. Bell\'etoile}
\author{D. Charrier}
\author{R. Dallier}
\author{T. Gousset}
\author{F. Haddad}
\author{J. Lamblin}
\author{P. Lautridou}
\author{O. Ravel}
\affiliation{SUBATECH, 4 rue Alfred Kastler, BP
  20722, F44307 Nantes cedex 3}
\author{A. Lecacheux}
\affiliation{LESIA, Observatoire de Paris - Section de Meudon, 5
  place Jules Janssen, F92195 Meudon cedex}
\author{L. Denis}
\affiliation{Observatoire de Paris - Station de radioastronomie,
  F18330 Nan\c{c}ay}
\author{P. Eschstruth}
\author{D. Monnier-Ragaigne}
\affiliation{LAL, Universit\'e Paris-Sud, B\^atiment 200, BP 34,
  F91898 Orsay cedex}

\date{\today}

\begin{abstract}
Firm evidence for a radio emission counterpart of cosmic ray air showers is presented. By the use of an antenna array set up in coincidence with ground particle detectors, we find a collection of events for which both time and arrival direction coincidences between particle and radio signals are observed. The counting rate corresponds to shower energies $\gtrsim 5\times 10^{16}$~eV. These results open overwhelming perspectives to complete existing detection methods for the observation of ultra high-energy cosmic rays. 
\end{abstract}


\maketitle 

The origin and the nature of ultra high-energy cosmic rays observed above $10^{19}$~eV is one of the most challenging questions in astroparticle physics~\cite{cronin}. It is also of major importance for the understanding of extremely energetic phenomena in the Universe. In the 1960's, radio emission associated with the development of extensive air showers was predicted and measured~\cite{ask,week}. A flurry of experimental investigations, carried out at that time with a wide variety of setups and in different frequency domains, have provided initial information about signals from $10^{17}$~eV cosmic rays~\cite{Allan}. Plagued by difficulties (poor reproductibility, atmospheric effects, technical limitations) efforts ceased in the late 1970's to the benefit of ground particle~\cite{agasa} and fluorescence detections~\cite{fly}. With the advent of cosmic ray research involving giant surface detectors as in the Auger experiment~\cite{auger}, the radio detection, having potentially 100 \% duty cycle and sensitive to the longitudinal development of the showers, is now reconsidered. In recent years, making use of up to date techniques in transient analysis, several groups have undertaken the task of reinvestigating the phenomenology of radio pulses~\cite{casa-mia,cascade-grande,rice} which are to a large extent a \textit{terra incognita}. Using our experiment CODALEMA (COsmic ray Detection Array with Logarithmic ElectroMagnetic Antennas), located at the Nan\c cay Radio Observatory, firm evidence for a radio emission counterpart of cosmic ray air showers is presented. From 60 days of running time, 18 events show both time and arrival-direction coincidences between particle and radio signals. This detection rate corresponds to shower energies $\gtrsim 5\times 10^{16}$~eV. 

\begin{figure}[h]
\begin{center}
\includegraphics[width=8.3cm]{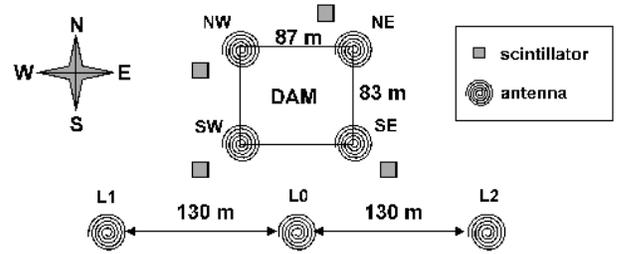}
\end{center}
\caption{CODALEMA setup for the second phase. The particle detector acts as a trigger with a fourfold coincidence requirement.}
\label{fig:setup2}
\end{figure}

During a first phase~\cite{rav04}, begun in March 2003, the trigger was generated using signals from one antenna devoted to this purpose. A good knowledge of the transient radio environment was acquired~\cite{dal03}. Coincident transients were measured from
 antennas spaced by several tens of meters, and information on the arrival direction and on the field amplitudes was obtained~\cite{dal04}. A first attempt to identify air shower candidate events was made based on gross features associated with air showers in the $10^{17}$~eV range~\cite{ard05}. However, due to the rather limited knowledge available concerning radio transients from the sky in the nanosecond range, the results obtained, although encouraging, could not be conclusive.

For the second phase~\cite{ard04}, to make sure that the measured transients were associated with air showers, four stations of particle scintillator detectors acting as a trigger have been added. This setup, including a new antenna configuration, is shown in Fig.~\ref{fig:setup2}.  It uses seven log-periodic antennas of the type constituting the Nan\c cay DecAMetric array (DAM)~\cite{DAM}. The antenna waveform signals are recorded after RF signal amplification (1-200 MHz, gain 35 dB) by LeCroy digital oscilloscopes (8-bit ADC, 500~MHz sampling frequency, 10~${\mu}$s recording time). To get enough sensitivity with these ADCs, the antennas are band pass filtered (24-82 MHz). 

The trigger corresponds to a fourfold coincidence within 600~ns from the particle detectors originally designed as prototype detectors for the Auger array~\cite{boratav}. Each 2.3~m$^2$ module (station) has two layers of acrylic scintillator, read out by a photomultiplier placed at the centre of each sheet. The photomultipliers have copper housings and it has been carefully verified that no correlation exists between high photomultiplier signal amplitudes and the presence of antenna signals. The signals from the upper layers of the four stations are digitized (8-bit ADC, 100~MHz sampling frequency, 10~${\mu}$s recording time). A station produces a signal when a coincidence between the two layers is obtained within a 60~ns time interval. This results in a counting rate of around 200~Hz. The rate of the fourfold coincidence of the
 four stations is about 0.7 event per minute. The stations are located close to the corners of the DAM array, and in this configuration the particle detectors delimit an active area of roughly~$7\times 10^3$~m$^2$. Using arrival times from the digitized photomultiplier signals, it is possible to determine the direction of the shower by triangulation with a plane fit. From the arrival direction distribution, a value of $16\times 10^3$~m$^2\times$sr is obtained for the acceptance, which corresponds to an energy threshold of about $1\times 10^{15}$~eV.

\begin{figure}
\begin{center}
\includegraphics[width=8.3cm]{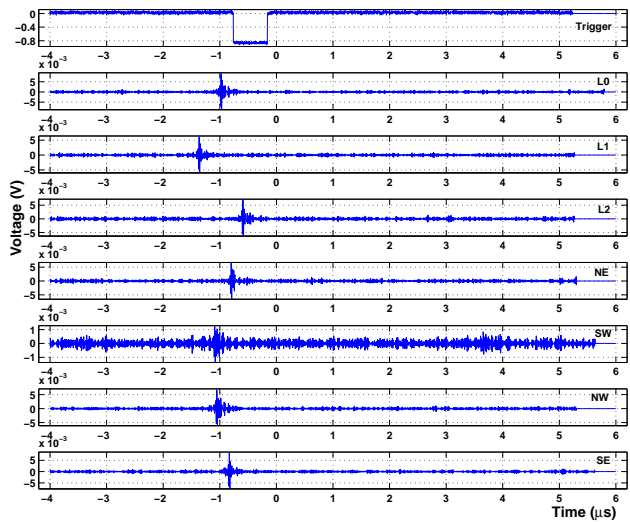}
\end{center}
\caption{Example of a seven antenna event, numerically filtered in the frequency band 37-70~MHz, obtained in coincidence with a particle trigger (first trace).}
\label{fig:Coinc.eps}
\end{figure}

For each fourfold coincidence from the particle detectors, the seven antenna signals are recorded. Due to the relatively low energy threshold, only a small fraction of these air shower events is expected to be accompanied by significant radio signals. 

The recognition of the radio transients is made during the offline analysis (see Fig.~\ref{fig:Coinc.eps}). For each waveform, the maximum value within the 37-70~MHz band is searched for in the time interval $[2~\mu\mathrm{s}~;~0~\mu\mathrm{s}]$. Then the noise level is determined using the mean ${\mu_n}$ and standard deviation ${\sigma_n}$ of the signal outside of a $\pm 1.2~\mu$s zone around this maximum. A pulse transient is flagged as "detected" when, in a $\pm 0.3~\mu$s zone around the maximum, the average of the squared signal ${\mu_s}$ satisfies the condition $\mu_s\ge \mu_n+k\times\sigma_n$ where the number $k$ is adjusted empirically to optimize the rejection of non pulse-like signals. 

For each antenna, if the threshold condition is fulfilled, the time of the maximum is determined. When at least 3 antennas are flagged it becomes possible to apply a triangulation procedure~\cite{ard05}, and the event is declared to be a radio candidate if the arrival direction obtained is plausible, \textit{i.e.} above the horizon. 

On average, there is one radio event for every $10^3$ particle triggers. This indicates that the energy threshold for radio detection is substantially higher than that of the particle array. 

\begin{figure}
\begin{center}
\includegraphics[width=8.3cm]{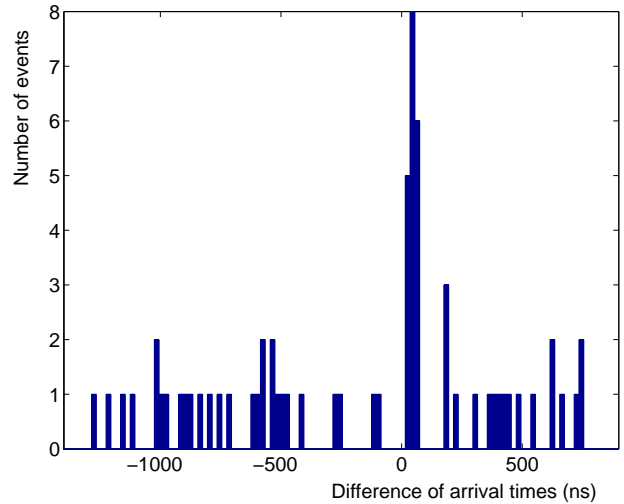}
\end{center}
\caption{Time delay between the radio plane front and the particle plane front} \label{fig:deltat.eps}
\end{figure}

Event selection then proceeds using the arrival times. The time of passage of the radio wavefront through a reference point is compared to the particle front time extracted from the scintillator signals. The distribution obtained is shown in Fig.~\ref{fig:deltat.eps}. The data correspond to 59.9 acquisition days and 70 antenna events. A very sharp peak (a few tens of nanoseconds) is obtained showing an unambiguous correlation between certain radio events and the particle triggers. This peak is centred around 40~ns with a systematic error of 20~ns. The delay between electric field and particle shower times could be measured with our apparatus, though more thorough studies and higher statistics will be necessary. As expected, in the 2~$\mu$s window where
 the search is conducted, radio transients can occur which are not associated with air showers and correspond to accidentals. Uncorrelated to the particles, such events have an uniform arrival time distribution.

\begin{figure}
\begin{center}
\includegraphics[height=8.3cm,angle=90]{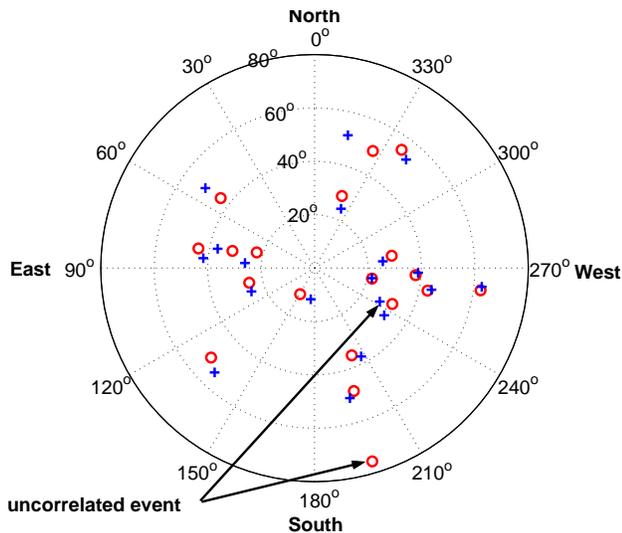}
\end{center}
\caption{Event arrival directions. The circles indicate directions
reconstructed from antenna signals whereas the crosses correspond to
directions given by the scintillators. Except for the marked event, each
circle is associated with the nearest cross.}
\label{fig: Event arrival direction}
\end{figure}

Finally, if the time-correlated events correspond to extensive air showers, the arrival directions reconstructed from  both scintillator and antenna data should be strongly correlated. Fig.~\ref{fig: Event arrival direction} exhibits, for the 19 events located in the main peak of Fig.~\ref{fig:deltat.eps}, the reconstructed arrival directions. The distribution of the angle between the two reconstructed directions is, as expected, a gaussian distribution centred on zero multiplied by a sin function coming from the solid angle factor. The width of the corresponding gaussian is about 3.5 degrees. The angle for the remaining uncorrelated event is much bigger, thus this event is certainly an accidental: its arrival direction given by the antennas is close to the horizon, which is typical of events from radio interference due to human activity. Moreover, the presence of one chance event in the peak is fully compatible with the observed uniform distribution in the 2~$\mu$s window.

For the 18 correlated events of Fig.~\ref{fig: Event arrival direction} the antenna multiplicity varies from 3 to 7 and the amplitude of the filtered signals goes up to $1.2$~mV$/$m. In the case of the seven-antenna event of Fig.~\ref{fig:Coinc.eps}, all the filtered signal amplitudes are around $0.25$~mV$/$m. This signal level corresponds to typical values expected for air shower energies in the $ 10^{17}$~eV range~\cite{Allan}. In order to obtain a rough estimation of the energy threshold for our setup, we suppose that the acceptance is the same for the two types of detectors (\textit{i.e.} $16\times 10^3$~m$^2\times$sr). The observed event rate then leads to an approximate energy threshold of $5\times 10^{16}$~eV.

In conclusion, these results strongly support the claim that electric field transients are generated by extensive air showers and that they can be clearly measured using wide band techniques. More data and technical upgrading are needed to examine the contribution that radio detection could bring in determining the energy and the nature of the cosmic rays. Various improvements are planned for CODALEMA: setting up of additional scintillators will make possible the determination of the shower energy and core position; extension of the W-E antenna line will enable to better sample the radio signal spread; increase of the ADC dynamics up to 12-bit encoding will allow to record the full 1~-~100~MHz frequency band, shower parameters could then be inferred from the full signal shape. In a subsequent upgrade, it is planned to install autonomous dipoles equipped with active front-end electronics, self-triggered and self-time-tagged. This is part of current investigations to use the potentialities of the radio detection techniques in conjunction with existing surface detectors. Among various particles capable to generate air showers, neutrinos are thought to be probably the only ones which could address the hadronic nature and source location of ultra energetic cosmic rays. Taking into account the ability of the radio detection method to characterize inclined extensive air showers~\cite{gousset}, the presented results could considerably enlarge the scope of such studies for cosmology related phenomena.

We thank P. Fleury for his continuous support.


\begin{thebibliography}{00}

\bibitem{cronin} J.W. Cronin, Rev. Mod. Phys. 71, S165-172 (1999).


\bibitem{ask} G.A. Askar'yan, Soviet Physics, J.E.T.P., 14, (2) 441 (1962) 

\bibitem{week} T.C. Weekes, Proc. of the first int. workshop on "Radiodetection of high Energy Particles", Los Angeles, November 16-18, 2000, AIP Conference Proceedings Vol 579 (2001) p. 3-13,

\bibitem{Allan} H.R. Allan, in: Progress in elementary particle and cosmic ray physics, ed. by J.G. Wilson and S.A. Wouthuysen (North Holland, 1971), p. 169.

\bibitem{agasa} N. Hayashida et al., Phys. Rev. Lett. 73, 3491 (1994); M. Takeda et al., Phys. Rev. Lett. 81, 1163 (1998).

\bibitem{fly} D.J. Bird et al., Phys. Rev. Lett. 71, 3401 (1993); Astro-phys. J. 441, 144 (1995).

\bibitem{auger} Auger Collaboration, Nucl. Instrum. Meth. A 523 (2004) 50-95 

\bibitem{casa-mia} K. Green, J.L. Rosner, D.A. Suprun, J.F. Wilkerson, Nucl. Instrum. Meth. A498, 256 (2003).

\bibitem{cascade-grande} A. Badea et al., Proceedings of CRIS2004, Nucl. Phys. Proc. Suppl. 136, 384 (2004); astro-ph/0409319.

\bibitem{rice} I. Kravchenko et al., astro-ph/0306408.

\bibitem{rav04} O. Ravel, R. Dallier, L. Denis, T. Gousset, F. Haddad, P. Lautridou, A. Lecacheux, E. Morteau, C. Rosolen, C. Roy, Proceedings of the 8th Pisa Meeting on Advanced Detectors "Frontiers Detectors for Frontier Physics", Nucl. Instr. Meth. A518, 213 (2004).

\bibitem{dal03} R. Dallier, L. Denis, T. Gousset, F. Haddad, P. Lautridou, A. Lecacheux, E. Morteau, O. Ravel, C. Rosolen, C. Roy, \textit{SF2A 2003 Scientific Highlights}, ed. F. Combes, \textit{et al.} (EDP Sciences, 2003).

\bibitem{dal04} A. Bell\'etoile, D. Ardouin, D. Charrier, R. Dallier, L. Denis, P. Eschstruth, T. Gousset, F. Haddad, P. Lautridou, A. Lecacheux, D. Monnier-Ragaigne, A. Rahmani, O. Ravel, \textit{SF2A 2004 Scientific Highlights}, ed. F. Combes \textit{et al.} (EDP Sciences, 2004), astro-ph/0409034.

\bibitem{ard05} D. Ardouin, A. Bell\'etoile, D. Charrier, R. Dallier, L. Denis, P. Eschstruth, T. Gousset, F. Haddad, J. Lamblin, P. Lautridou, A. Lecacheux, D. Monnier-Ragaigne, A. Rahmani, O. Ravel, to be submitted to Nucl. Instrum. Meth. A. 

\bibitem{ard04} D. Ardouin, A. Bell\'etoile, D. Charrier, R. Dallier, L. Denis, P. Eschstruth, T. Gousset, F. Haddad, J. Lamblin, P. Lautridou, A. Lecacheux, D. Monnier-Ragaigne, A. Rahmani, O. Ravel, Proceedings of the 19th European Cosmic Ray Symposium, Florence, 2004, astro-ph/0412211.

\bibitem{DAM} http://www.obs-nancay.fr/  (2005)

\bibitem{boratav} M. Boratav, J.W. Cronin, B. Dudelzak, P. Eschstruth, P. Roy, V. Sahakian and Z. Strachman, Proceedings of the 24th ICRC, Rome, 954(1995).

\bibitem{gousset} T. Gousset, O. Ravel and C. Roy, Astropart. Phys. 22, 103-107 (2004).

\end{thebibliography}
\end{document}